\documentclass[twocolumn,showpacs,preprintnumbers,amsmath,amssymb,aps,prb]{revtex4}

\usepackage{graphicx}
\usepackage{graphics}
\usepackage{dcolumn}
\usepackage{bm}
\usepackage{color}

\begin{document}

\title{Effects of Andreev reflection on the conductance of quantum-chaotic dots}

\author{Victor A. Gopar}
\affiliation{Departamento de F\'isica Te\'orica and Instituto de Biocomputaci\'on y F\'isica de Sistemas Complejos (BIFI), Universidad de Zaragoza, Pedro Cerbuna 12, E-50009 Zaragoza, Spain}

\author{J. A. M\'endez-Berm\'udez}
\affiliation{Instituto de F\'{\i}sica, Universidad Aut\'onoma de Puebla, Apartado Postal J-48, Puebla 72570 Mexico}

\author{Arafa H. Aly}
\affiliation{Departamento de F\'isica Te\'orica and Instituto de Biocomputaci\'on y F\'isica de Sistemas Complejos (BIFI), Universidad de Zaragoza, Pedro Cerbuna 12, E-50009 Zaragoza,  Spain}
\affiliation{Physics Department, Faculty of Sciences, Beni-Suef University, Egypt}


\begin{abstract}
We investigate the conductance statistics of a quantum-chaotic dot---a normal-metal grain---with a superconducting lead attached to it. The cases of one and two normal leads additionally attached to the dot are studied. For these two configurations the complete distribution of the conductance is calculated, within the framework of random matrix theory, as a function of the transparency parameter of the Schottky barrier formed at the interface of the normal-metal and superconducting regions. Our predictions are verified by numerical simulations.
\end{abstract}


\maketitle

\section{Introduction}

The physics of normal-metal--superconducting (NS) hybrid structures has
been of interest for several decades. In the sixties, the scattering of a quasi-particle at a NS interface was first described by Andreev. \cite{andreev} Thus, it is now widely known that an electron approaching to a superconductor from the normal-metal side, with an energy above the Fermi level, can be reflected at the interface as a hole with an energy below the Fermi level and  same momentum, but in opposite direction. A Cooper pair is thus formed in the superconductor. This scattering process is known as Andreev reflection.
More recently, the interest in this subject was renewed in the field of mesoscopic physics, where the electron wave function remains coherent and therefore quantum interference effects play an important role. Several effects of phase-coherence Andreev scattering are under current investigation. In addition, the effects of disorder and ballistic-chaotic scattering in hybrid structures have been widely studied. For example quantities such as the density of states and the Ehrenfest time as well as phenomena such as coherent backscattering, excitation and gap fluctuations, superconducting proximity effect, and resonant tunneling have been investigated in mesoscopic hybrid structures.\cite{lambert_1, kosztin, asano, lodder, altland, adagideli, vavilov,  beenakker_1, kormanyos, kormanyos97, tero, samuelsson1,samuelsson, goorden} Also, the effects of Andreev scattering on the conductance through quantum dots whose classical dynamics is integrable or mixed (regular and chaotic) have been investigated.\cite{clerk,nikolaos}

On the other hand, random matrix theory (RMT) has been useful to study the statistical properties of many different phenomena in normal metals \cite{beenakker_review, pier_book, bulgakov, gopar_1, gopar_2} as well as in superconductors. In particular, the effects of NS interfaces on the conductance of disordered and chaotic billiards--quantum dots--have received considerable attention.\cite{melsen, vavilov,  beenakker_1, kormanyos, tero, beenakker_2, brouwer_1, slevin, brouwer_2, beri}
Recently, some statistical properties of the conductance fluctuations in mesoscopic NS heterostructures have been experimentally studied in Ref.~[\onlinecite{jespersen}] where the  explanation of the experimental results was based on RMT of quantum transport through chaotic quantum dots; it was also pointed out the importance of considering the potential barriers formed at the NS interface.

Using a scattering approach to the problem of quantum transport and RMT, here we study the statistical properties of the conductance through a quantum-chaotic dot with a superconducting lead attached to it.
In particular, we are interested in the distribution of the conductance.
The configurations of one and two normal leads additionally attached to the dot are considered. We take into account the presence of a Schottky barrier--which is formed in real NS junctions-- of transparency $\Gamma$ in the superconducting lead. Thus, for these two setups we calculate the complete distribution of conductances for different values of $\Gamma$. As we will see, the conductance distribution is affected by the presence of the superconducting region as well as the scattering at the Schottky barrier.

This paper is organized as follows. In the next Section we briefly review general ideas about electron-hole scattering at NS junctions. We also introduce the scattering matrices associated to the different elements of our experimental setup.
In Sec. III, we study the two-terminal configuration. Firstly, we calculate the total scattering matrix and the conductance. Secondly, we present our results for the distribution of the conductance for the cases of time-reversal symmetry ($\beta=1$) and broken time-reversal symmetry ($\beta=2$). Simple analytical expressions are provided for the conductance distribution. Sec. IV is devoted to the three-terminal configuration. We calculate the necessary scattering-matrix elements in order to calculate the conductance and present the results for its distribution for both symmetries ($\beta=1$ and $\beta=2$), at different values of the transparency of the Schottky barrier. A summary and conclusions of our work are presented in Sec. V.

\section{Model and formalism}

\begin{figure}
\includegraphics[width=0.65\columnwidth]{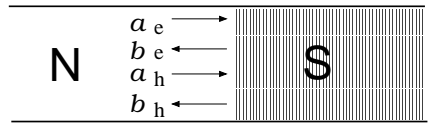}
\caption{Illustration of a normal-metal--superconductor (NS) junction.
The coefficients $(a_e,a_h)$ and $(b_e,b_h)$ are related by the scattering matrix $S^A$, see Eq.~(\ref{matrixSA}).}
\label{fig_1}
\end{figure}

We start our study with some general ideas on the scattering mechanism of electrons in a NS junction, as illustrated in Fig.~\ref{fig_1}. As we mentioned above, in a NS junction an incident electron from the normal side can be reflected by the superconductor as an electron or as a hole. The electron and hole wave functions $\psi^e$ and $\psi^h$, respectively, satisfy the Bogoliubov-De Gennes equation\cite{gennes} given by
\begin{equation}
\label{BdeG}
\left(
\begin{array}{cc}
H_0 &  \Delta   \\
\Delta^* & -H^*_0
\end{array}
\right)\;\left( \begin{array}{c}
           \psi^e \\ \psi^h
         \end{array} \right)
 = \epsilon \left(\begin{array}{c}
           \psi^e \\ \psi^h
         \end{array} \right) \ ,
\end{equation}
where $H_0$ is the single electron Hamiltonian and $\Delta$ is the potential which couples the electron and hole wave functions. We assume that $H_0$ is independent of spin-orbit interactions. On the other hand, far from the NS boundary the superconducting gap $\Delta$ goes to zero in the normal-metal region, whereas in the superconducting region the pair potential has an amplitude $\Delta_0$ and phase $\phi$. Near the NS interface, however, $\Delta$ depends on the details of the NS junction; assuming that the width of the NS interface is small compared to the superconducting coherence length and neglecting any pairing interaction in the normal region, the pair potential is usually modeled as a step potential: $\Delta(r)=\Delta_0 e^{i \phi} \Theta(x)$, $\Theta(x)$ being the step function. Now we consider that the energy of an incident electron is in the energy gap ($E_F < E < E_F+\Delta_0$, $E_F$ being the Fermi energy) where there are no propagating wave solutions in the superconducting region, i.e., the electron is reflected as a hole or electron. Let us denote by $r^A_{he}$ the reflection matrix describing the Andreev reflection. Within the Andreev approximation, where only Andreev scattering processes are taking into account and with the above assumptions for $\Delta$, a solution of Eq.~(\ref{BdeG}) in the normal region can be written as\cite{slevin, beenakker_review}
\begin{eqnarray}
\label{solutionBdeG}
\left(
\begin{array}{c}
\psi^e  \\ \psi^h
\end{array}
\right)\;&=&\left( \begin{array}{c}
           1 \\ 0
         \end{array} \right)\exp[{ik^e_nx}]\phi_{n}(y,z,E_F+\epsilon) \nonumber
         \\ &+&\left(\begin{array}{c}
           0 \\ r^A_{he}
         \end{array} \right)\exp[{ik_n^hx}]\phi_{n}(y,z,E_F-\epsilon) \ .
\end{eqnarray}

In this expansion, $\phi_{n}(y,z)$ is the transverse wave function that carries a current in the  $+x$ direction and $n$ labels the transverse mode--open channel--, while the wave number $k_n^{e, h}$ is given by $k_n^{e}=\sqrt{2m/\hbar^2}\sqrt{E_F-E_n + \epsilon}$ and $k_n^{h}=\sqrt{2m/\hbar^2}\sqrt{E_F-E_n -\epsilon}$.

In the linear-response limit, i.e., zero bias voltage ($\epsilon= 0$), we have  $k_n^{e}=k_n^{h}$.
In this limit, using the continuity of the wave function and its derivative
at the interface, the reflection matrix $ r^A_{he}$ is given by  $r^A_{he}=i\exp{(-i\phi)}{\bf1}_N$ and $r^A_{eh}=i\exp{(i\phi)}{\bf1}_N$, where ${\bf1}_N$ is the $N\times N$ unit matrix.

\subsection{Andreev scattering matrix}

As we have mentioned, we consider attached leads supporting one open channel, therefore from now on we specialize to this case. Thus, the $2\times2$ scattering matrix for the Andreev reflection which relates the coefficients $(a_e,a_h)$ and $(b_e,b_h)$, see Fig.~\ref{fig_1}, is given by
\begin{equation}
\left( \begin{array}{c}
           b_e \\ b_h
         \end{array} \right)=
S^A \left( \begin{array}{c}
           a_e \\ a_h
         \end{array} \right) \ ,
\end{equation}
with
\begin{equation}
\label{matrixSA}
S^A=
\left(
\begin{array}{cc}
0 &  r^A_{eh}   \\
r^A_{he} & 0
\end{array}
\right) \ ,
\end{equation}
where $r^A_{he}=i\exp{(-i\phi)}$ and $r^A_{eh}=-(r^{A}_{he})^*$.


\subsection{Schottky barrier}

We also consider the potential barrier created at the NS interface
due to the mismatch of the conduction bands of the two different regions.
We model this Schottky barrier as a planar barrier characterized by its transparency $\Gamma$. It is of relevance to consider the effect of this potential on the transport since, as we will see later, the conductance and therefore its statistical properties are affected depending on the strength of $\Gamma$.

On the other hand, since in the normal region electrons and holes are not coupled, the scattering matrix of the Schottky barrier $S^S$ can be written as
\begin{equation}
\label{matrixSS0}
S^{S}=
\left(
\begin{array}{cc}
S^S_e &  0   \\
0 & S_h^S
\end{array}
\right) \ ,
\end{equation}
with the scattering matrix for electrons $S^S_e$ given by\cite{brouwer_3, moises}
\begin{equation}
\label{matrixSS}
S_e^{S}=
\left(
\begin{array}{cc}
\sqrt{1-\Gamma} &  i\sqrt{\Gamma}   \\
i\sqrt{\Gamma} & \sqrt{1-\Gamma}
\end{array}
\right) \ ,
\end{equation}
while the scattering matrix for holes $S_h^{S}$ satisfies $S_h^{S}=(S_e^{S})^*$. \cite{slevin, beenakker_review}


\subsection{Scattering matrix of the quantum dot}
\label{introductionC}

We now introduce the polar representation for the scattering matrix $S^D$ associated to the quantum dot. For electrons, the  $2\times 2$ $S_e^D$-matrix is written as
\begin{equation}
\label{polar}
S_e^{D}=
\left(
\begin{array}{cc}
-\sqrt{1-\tau}e^{i(\theta+\theta')} &  \sqrt{\tau} e^{i(\theta+\psi')}  \\
\sqrt{\tau}e^{i(\theta'+\psi)} & \sqrt{1-\tau}e^{i(\psi+\psi')}
\end{array}
\right) \ ,
\end{equation}
where $\tau$ and the phases $\theta$, $\theta'$, $\psi$, and $\psi'$ are
independent variables.
In presence of time reversal symmetry, $S_e^D$ is a symmetric matrix and
therefore $\theta=\theta'$ and $\psi=\psi'$.
As before, since in the normal region the electrons and holes are not coupled, the complete scattering matrix for the quantum dot can be written as
\begin{equation}
\label{matrixSD}
S^{D}=
\left(
\begin{array}{cc}
S_e^D &  0  \\
0 & S_h^D
\end{array}
\right) \ ,
\end{equation}
with $S_h^D=(S_e^D)^*$.


\section{Two-terminal configuration}

First we consider a two-terminal setup. As shown in Fig.~\ref{fig_2}, a normal lead is attached to a normal quantum dot while a second lead with a superconducting region and a Schottky barrier of transparency $\Gamma$ is also attached.
Both leads support one open channel. In a similar two-terminal setup and for a large number of channels in the leads, the average and variance of the conductance have been studied in Ref.~[\onlinecite{clerk}].

\subsection{Total scattering matrix and conductance}

\begin{figure}
\includegraphics[width=0.65\columnwidth]{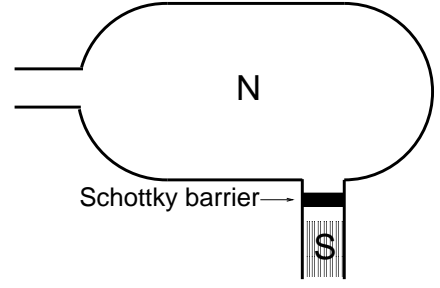}
\caption{Two-terminal setup. A normal-metal lead (left) and a superconducting lead (down) are attached to a quantum dot. Each lead supports one open channel. A Schottky barrier of transparency $\Gamma$ is introduced in the superconducting lead.}
\label{fig_2}
\end{figure}

We are ready to calculate the full scattering matrix of the system, Fig.~\ref{fig_2}, since we have introduced already the scattering matrices associated to the components of our hybrid structure in the previous section, i.e., we only need to combine properly the scattering matrices $S^A$, $S^S$, and $S^D$ from Eqs.~(\ref{matrixSA}), (\ref{matrixSS0}), and (\ref{matrixSD}), respectively.

We combine first the scattering matrices associated to the Andreev reflection $S^A$ and Schottky barrier $S^S$. The resulting scattering matrix $S^{SA}$ is given by
\begin{equation}
 S^{SA}=r^S+t'^{S}S^A\left(1-r'^S S^A \right)^{-1}t^S,
\end{equation}
where we have defined the matrices
\begin{equation}
 r^S=\left(
\begin{array}{cc}
r_e^S &  0  \\
0 & r_h^S
\end{array}
\right)\;, \ \  t^S=\left(
\begin{array}{cc}
t_e^S &  0  \\
0 & t_h^S
\end{array}
\right)\;.
\end{equation}
From Eq.~(\ref{matrixSS}), $r_e^S=r_h^S=\sqrt{1-\Gamma}$, $t_e^S=i\sqrt{\Gamma}$, and $t_h^S=-i\sqrt{\Gamma}$, we obtain
\begin{eqnarray}
\label{smatrixSA}
 &&S^{SA} = \nonumber \\
&&\frac{1}{\Delta}\left(
\begin{array}{cc}
\Delta r_e^S +t_e'^{S} r^A_{eh} r_h'^S r'^A_{he} t_e^S  & t_e'^{S} r^A_{eh} t_h^S \\ \\
t_h'^{S} r^A_{he} t_e^S  & \Delta r_h^S +t_h'^{S} r^A_{he} r_e'^S r'^A_{eh} t_h^S \\
\end{array}\right)\;, \nonumber \\
\end{eqnarray}
with $\Delta=1-r_h'^S r_{he}^A r_e'^S r_{eh}^A$. We now combine the above expression for
$S^{SA}$ with the scattering matrix of the quantum dot $S^D$.
The resulting scattering matrix $S$ is given by
\begin{equation}
 S=r^D+t'^{D}S^{SB}\left(1-r'^D S^{SB} \right)^{-1}t^D.
\end{equation}
After some algebra we write the total matrix $S$ as
 \begin{equation}
\label{totalsmatrix}
 S  =  \left(
\begin{array}{cc}
s_{ee}  & s_{eh} \\
s_{he} & s_{hh}
\end{array}\right) \ ,
\end{equation}
where
\begin{eqnarray}
\label{elementsofstotal}
 s_{ee} &=& 2Ce^{2i(\theta+\theta')}\big[(\Gamma-2)\sqrt{1-\tau}+ \nonumber \\
& & \sqrt{1-\Gamma}
\left((2-\tau)\cos{(\psi+\psi')}+i\tau \sin{(\psi+\psi')}\right) \big] \ , \nonumber \\
s_{eh}  &=& -iCe^{i\left(\phi+\psi-\psi'+\theta-\theta'\right)} \Gamma \tau \ , \nonumber \\
s_{he} &=& -iCe^{-i\left(\phi+\psi-\psi'+\theta-\theta'\right)} \Gamma \tau \ , \nonumber \\
s_{hh} &=& 2Ce^{-2i(\theta+\theta')}\big[(\Gamma-2)\sqrt{1-\tau}+ \nonumber  \\ & & \sqrt{1-\Gamma}\left((2-\tau)
\cos{(\psi+\psi')}-i\tau \sin{(\psi+\psi')}\right) \big] \ , \nonumber \\
\end{eqnarray}
with
\[
C=[(\Gamma-2)(\tau-2)-4\sqrt{(\Gamma-1)(\tau-1)}\cos(\psi+\psi')]^{-1} \ .
\]

Finally, the dimensionless conductance $g$ due to the presence of the NS interface
in our two-terminal setup is given, in terms of elements of the total scattering matrix, by\cite{blonder, lambert_1, takane}
\begin{equation}
\label{conductanceNS}
 g=\frac{G}{G_0}=2 s_{he} s_{he}^{\dagger}  \ ,
\end{equation}
where $G_0 (=2e^2/h)$ is the conductance quantum.

\subsection{Conductance distribution}

We now calculate the distribution of the conductance by assuming that the statistics of the scattering matrix of the quantum dot $S^D$ follows one of the circular ensembles, namely, the circular orthogonal ensemble ($\beta=1$) or the circular unitary ensemble ($\beta=2$).

From Eq.~(\ref{conductanceNS}), the distribution of the conductance is given by
\begin{equation}
\label{defpofg}
P^{(\beta)}(g)=\int \delta \left(g-2 s_{he} s_{he}^{\dagger}  \right)d\mu^{(\beta)}(S^D) \ ,
\end{equation}
with invariant measure\cite{pier_book}
\begin{equation}
\label{dmu}
 d\mu^{(\beta)}(S^D)=\left\{
\begin{array}{ll}
{\displaystyle \frac{d\tau}{2\sqrt{\tau}}\frac{d\theta}{2\pi}\frac{d\psi}{2\pi} } & \mathrm{for} \ \ \beta=1\\
\\
{\displaystyle d\tau\frac{d\theta}{2\pi}\frac{d\psi}{2\pi}\frac{d\theta'}{2\pi}\frac{d\psi'}{2\pi} } & \mathrm{for} \ \ \beta=2
\end{array} \right. \ .
\end{equation}

In the following subsections we provide analytical expressions for the distribution of the conductance for the two-terminal configuration by the evaluation of the integrals in Eq.~(\ref{defpofg}), for the cases $\beta=1$ and $\beta=2$.


\subsubsection{Time-reversal-symmetric case, $\beta=1$}

We recall that for the orthogonal symmetry, $S^D_e$ and $S^D_h$ are symmetric scattering matrices and therefore $\theta=\theta'$ and $\psi=\psi'$ in the polar representation of Eq.~(\ref{polar}).

Thus, from Eq.~(\ref{elementsofstotal}) we find
\begin{equation}
\label{gbeta1}
s_{he} s_{he}^{\dagger}=\frac{\Gamma^2 \tau^2}{\left[(\Gamma-2)(\tau-2)-4\sqrt{(\Gamma-1)(\tau-1)}\cos(2\psi)\right]^2} \ .
\end{equation}
Before considering the general case of arbitrary $\Gamma$ let us consider the particular case $\Gamma=1$, i.e., maximum transparency of the Schottky barrier. In this case Eq.~(\ref{gbeta1}) reduces to
\begin{equation}
 s_{he} s_{he}^{\dagger}=\frac{\tau^2}{(2-\tau)^2} , {\hspace{1cm} \mathrm{for} \ \ \Gamma=1} \ .
\end{equation}
Thus from Eqs.~(\ref{conductanceNS}) and (\ref{defpofg}), we obtain the simple expression for the distribution
\begin{equation}
\label{pofgamma1}
 P^{(1)}(g)=\frac{1}{2\left[\sqrt{g}\left(\sqrt{2}+\sqrt{g}\right) \right]^{3/2}} \ , \ \ \mathrm{for} \ \ \Gamma=1,
\end{equation}
which we have already normalized.

\begin{figure}
\includegraphics[width=0.95\columnwidth]{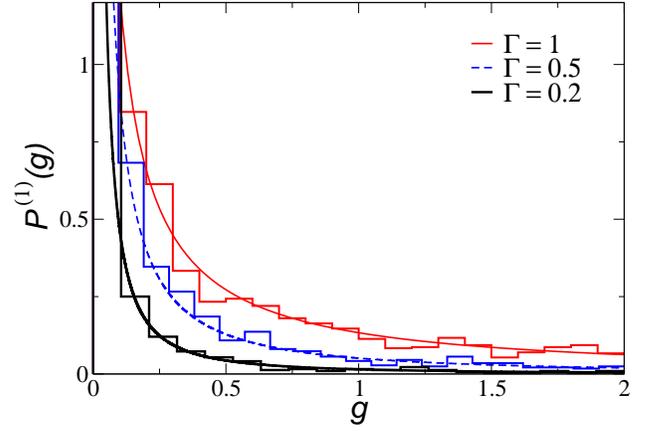}
\caption{(Color online) Distribution of the conductance $P^{(1)}(g)$ for the two-terminal setup for different values of the transparency $\Gamma$. Good agreement is observed between theory [Eqs.~(\ref{pofgamma1}) and (\ref{pofggeneralbeta1})] and the corresponding numerical simulations (histograms).}
\label{fig_3}
\end{figure}

We now study the case of arbitrary $\Gamma$. Therefore
we need to perform the integrals in Eq.~(\ref{defpofg}) with $s_{he} s_{he}^{\dagger}$ given by Eq.~(\ref{gbeta1}). The calculation is direct but somewhat lengthy. We finally get
\begin{eqnarray}
\label{pofggeneralbeta1}
 P^{(1)}(g)&=&\frac{\pi^2}{\Gamma}\frac{1}{g^{3/2}}\left[\frac{\sqrt{B^2-4AC}-B}{C^3}\right]^{1/2} \times \nonumber  \\ && \ \ \ \ \ \ \ \  \ \  \  \left[E(x,m)-F(x,m) \right] \Big\vert_{\tau_-}^{\tau_+} \ ,
\end{eqnarray}
where
\begin{eqnarray*}
A&=&16(1-\Gamma)-4(\Gamma-2)^2 \ , \\
B&=& 4(\Gamma-2)^2+16(1-\Gamma+4\sqrt{\frac{2}{g}}\left(2-\Gamma\right)\Gamma \ , \\
C&=&2\sqrt{\frac{2}{g}}\Gamma(\Gamma-2)-(\Gamma-2)^2-\frac{2\Gamma^2}{g} \ , \\
m&=& \frac{B+\sqrt{B^2-4AC}}{B-\sqrt{B^2-4AC}} \ , \\
x&=&i\mathrm{arcsinh}\left(\sqrt{\frac{2C\tau}{B+\sqrt{B^2-4AC}}} \right) \ .
\end{eqnarray*}
$F(x,m)$ and $E(x,m)$ in Eq.~(\ref{pofggeneralbeta1}) are the elliptical integrals of first and second kind, respectively, which have to be evaluated at the limits $\tau_+$ and $\tau_-$ given by
\begin{equation}
 \tau_{\pm}=\left[ \frac{1}{2}+\left(\frac{\sqrt{2}-\Gamma}{\sqrt{2\Gamma}}\right)\frac{1}{\sqrt{g}} \mp \frac{1}{\Gamma}\sqrt{\frac{(2-g)(1-\Gamma)}{g}} \right]^{-1} \ .
\end{equation}
We have verified our predictions for $P^{(1)}(g)$, Eqs.~(\ref{pofgamma1})
and (\ref{pofggeneralbeta1}), by comparing with numerical simulations of a
quantum-chaotic dot (in all our numerical simulations we have considered a Bunimovich stadium geometry whose classical dynamics is chaotic). In Fig.~\ref{fig_3}, we show both numerical and analytical results for $\Gamma=0.2$, 0.5, and 1.
We observe good agreement between theory and the corresponding numerical simulations.

\begin{figure}
\includegraphics[width=0.95\columnwidth]{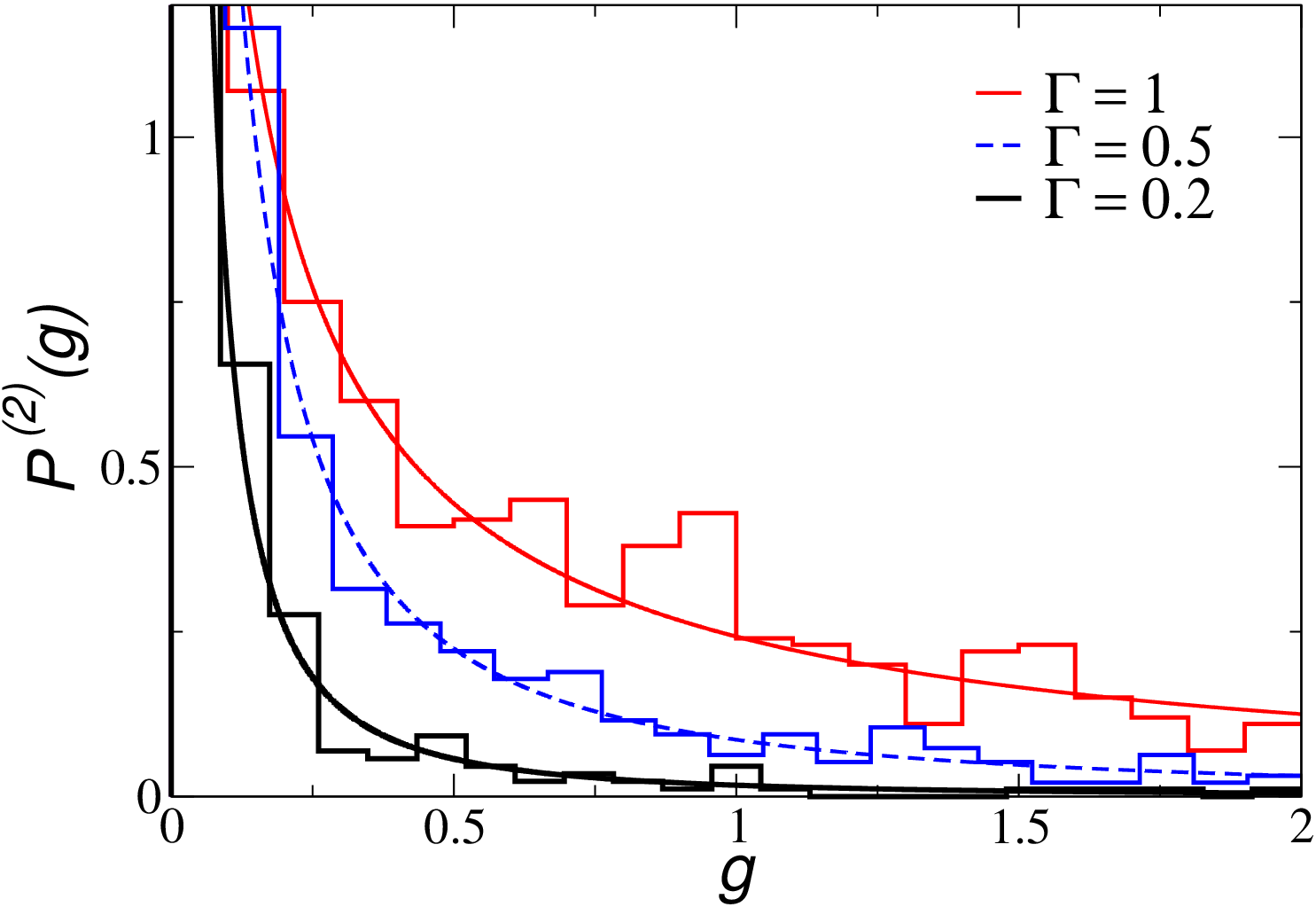}
\caption{(Color online) Distribution of the conductance $P^{(2)}(g)$ for the two-terminal setup for different values of the transparency $\Gamma$. Good agreement is observed between theory [Eqs.~(\ref{pofggamma1beta2}) and (\ref{pofggeneralbeta2})] and the corresponding numerical simulations (histograms).}
\label{fig_4}
\end{figure}


\subsubsection{Broken time-reversal symmetry, $\beta=2$}

We now consider the circular unitary ensemble. We recall that this symmetry class is relevant for situations where time-reversal symmetry is absent. From Eq.~(\ref{elementsofstotal}) we have
\begin{eqnarray}
\label{gbeta2}
&& \hspace{-0.25cm} s_{he} s_{he}^{\dagger} \nonumber \\
&& = \displaystyle
\frac{2\Gamma^2 \tau^2}{\left[ (\Gamma-2)(\tau-2)-4\sqrt{(\Gamma-1)(\tau-1)}\cos(\psi+\psi')\right]^2} \ . \nonumber \\
\end{eqnarray}
As in the previous subsection we start with the special case $\Gamma=1$. Thus, from Eqs.~(\ref{conductanceNS}), (\ref{defpofg}), and (\ref{gbeta2}) the distribution of conductance is given by
\begin{equation}
\label{pofggamma1beta2}
 P^{(2)}(g)=\sqrt{\frac{2}{g}}\frac{1}{\left(\sqrt{2}+\sqrt{g}\right)^2} \ , \ \  \mathrm{for} \ \ \Gamma=1.
\end{equation}
The general case of arbitrary $\Gamma$ involves a direct but lengthy calculation, as in the $\beta=1$ case. For $\beta=2$, however, the distribution gets a simpler expression. After some algebra we find
\begin{equation}
\label{pofggeneralbeta2}
P^{(2)}(g)=\frac{\left[\Gamma(\sqrt{2g}-2)+4 \right]\Gamma^2}{\sqrt{g}\left[ \sqrt{2}\Gamma+\sqrt{g}\left(2-\Gamma \right)\right]^3} \ .
\end{equation}
In Fig.~\ref{fig_4} we plot $P^{(2)}(g)$, as given by Eqs.~(\ref{pofggamma1beta2}) and (\ref{pofggeneralbeta2}), and compare with numerical simulations of a quantum-chaotic dot. We choose $\Gamma=0.2$, 0.5, and 1. As for the case $\beta=1$, for broken time-reversal symmetry a good agreement between theory and numerics is also observed.

To conclude this Section we notice that, for the two-terminal configuration studied here and for both symmetries ($\beta=1$ and $\beta=2$), the distribution of the conductance becomes broader as the value of the transparency of the Schottky barrier is increased, see Figs.~\ref{fig_3} and \ref{fig_4}. This broadening of the distribution implies that large values of $g$ (values of $g$ at the tail of the distribution) become statistically favored as the Schottky barrier becomes transparent. We anticipate that the effect of the Schottky barrier on the conductance is somewhat different in the three-terminal configuration that we study in the following Section.


\section{Three-terminal configuration}

\begin{figure}
\includegraphics[width=0.65\columnwidth]{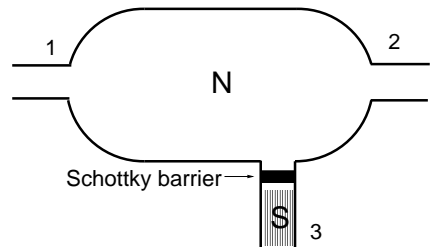}
\caption{Three-terminal setup. Two normal-metal leads (1 and 2) and a superconducting lead (3) are attached to a quantum dot. Each lead supports one open channel. A Schottky barrier of transparency $\Gamma$ is introduced in the superconducting lead.}
\label{fig_5}
\end{figure}

We now consider a more complex situation. As shown in Fig.~\ref{fig_5}, two normal leads are attached to a normal quantum dot while a third lead with a superconducting material and a Schottky barrier is also attached. All the leads support one open channel. As in the previous Section, we calculate the total scattering matrix of the hybrid structure. The present case is, however, algebraically more involved than the two-terminal configuration.

\subsection{Total scattering matrix and conductance}

We define a scattering matrix $S$ that relates the incoming and outgoing wave amplitudes in the normal leads (leads 1 and 2 in Fig.~\ref{fig_5}):
\begin{equation}
\left( \begin{array}{c}
           b_e^1 \\ b_e^2 \\ b_h^1 \\b_h^2
         \end{array} \right)=
S \left( \begin{array}{c}
           a_e^1 \\ a_e^2 \\a_h^1 \\ a_h^2
         \end{array} \right) \ ,
\end{equation}
where $S$ has the following structure
\begin{equation}
\label{matrixST2leads}
S =
\left(
\begin{array}{cccc}
s_{11}^{ee} &  s_{12}^{ee}  & s_{11}^{eh} & s_{12}^{eh} \\
s_{21}^{ee} &  s_{22}^{ee}  & s_{21}^{eh} & s_{22}^{eh} \\
s_{11}^{he} &  s_{12}^{he}  & s_{11}^{hh} & s_{12}^{hh} \\
s_{21}^{he} &  s_{22}^{he}  & s_{21}^{hh} & s_{22}^{hh}
\end{array}
\right) \ .
\end{equation}
Here, the subscripts denote the leads. See Fig.~\ref{fig_5}. We recall that we are working at energies where there are no propagating waves in the superconducting region.
In order to obtain the explicit expressions for the elements of the scattering matrix $S$, we need to combine the scattering matrix of the superconducting region [Eq.~(\ref{matrixSA})], the Schottky barrier [Eqs.~(\ref{matrixSS0}) and (\ref{matrixSS})], and the quantum dot. The scattering matrix associated to the quantum dot $S^D$ can be written in two blocks, as in Section \ref{introductionC}, one for the electrons and the other for the holes. Thus, for the three-terminal setup, we write for electrons
\begin{equation}
\label{matrixSDT2leads}
S^D_e =
\left(
\begin{array}{ccc}
r_{11}^{e} &  t_{12}^{e}  & t_{13}^{e}  \\
t_{21}^{e} &  r_{22}^{e}  & t_{21}^{e}  \\
t_{31}^{e} &  t_{32}^{e}  & r_{33}^{e}
\end{array}
\right) \ ,
\end{equation}
with a similar expression for the holes. Once we have introduced the scattering
matrices associated to the different parts of our setup,  we can follow the same steps of the previous Section to obtain the total scattering matrix.
On the other hand, however, the dimensionless conductance $g$ for our three-terminal setup is given by\cite{lambert_2}
\begin{equation}
\label{gtwoterminal}
g=\vert s_{21}^{ee} \vert ^2 + \vert s_{21}^{eh} \vert^2 \ .
\end{equation}
Thus, we need only the elements $s_{21}^{ee}$ and $s_{21}^{eh}$ of the total scattering matrix $S$, Eq.~(\ref{matrixST2leads}), in order to calculate the conductance. After some algebra we find for these two scattering-matrix elements
\begin{eqnarray}
\label{s21ee}
s_{21}^{ee}&=&A_2^e+\frac{1}{\Delta}C_2^e r_{eh}^A C^h r_{he}^A A^e \ , \\
\label{s21eh}
s_{21}^{eh} &=& \frac{1}{\Delta}C_2^h r_{he}^A A^e \ ,
\end{eqnarray}
where
\begin{eqnarray*}
\label{definitions}
\Delta&=& 1-r_{he}^A C^e r_{eh}^A C^h \ , \\
A_2^e&=&t_{21}^e+\frac{t_{23}^e r'^S_e t_{31}^e}{1-r_{33}^e r'^S_e} \ , \\
C_2^{e}&=& t_{23}^e \left[ t^S_e +\frac{r'^S_e r_{33}^e t^S_e}{1-r_{33}^e r'^S_e} \right] \ , \\
A^e&=& t'^S_e t_{31}^e\frac{1}{1-r_{33}^h r'^S_h} \ , \\
\label{definitions1}
C^e&=& r^S_e+\frac{t'^S_e r_{33}^e t^S_e}{1-r_{33}^e r'^S_e} \ .
\end{eqnarray*}
$C^h$ and $C_2^{h}$ in Eqs.~(\ref{s21ee}) and (\ref{s21eh}) are defined as $C^e$ and $C_2^{e}$, respectively, by replacing $e \to h$.
We can observe that if the third lead were absent, the expression for
conductance, Eq.~(\ref{gtwoterminal}), is reduced to  the usual Landauer-B\"uttiker formula $g=|t_{21}^e|^2\equiv|t_{21}|^2$.

For the particular values of $\Gamma =0$ and 1, the expressions for the elements $s_{21}^{ee}$ and $s_{21}^{eh}$ are simplified. Therefore for these two special values of $\Gamma$, it is useful to show the expressions of $s_{21}^{ee}$ and $s_{21}^{eh}$ since they exhibit explicitly the different scattering processes among the three leads attached to the quantum dot that contribute to the conductance.

\subsubsection{Case $\Gamma=1$}

For the case of zero reflection at the Schottky barrier the expressions for $s_{21}^{ee}$ and $s_{21}^{eh}$ in Eqs.~(\ref{s21ee}) and (\ref{s21eh}), respectively, are simplified to
\begin{eqnarray}
s_{21}^{ee}&=& t_{21}^e-\frac{t_{23}^e r_{33}^h t_{31}^e}{1+r_{33}^e r_{33}^h} \ , \\
s_{21}^{eh}&=& \frac{t_{23}^h t_{31}^e}{1+r_{33}^e r_{33}^h} \ .
\end{eqnarray}
Therefore the conductance is given by
\begin{equation}
g = \left\vert t_{21}^e-\frac{t_{23}^e r_{33}^h t_{31}^e}{1+r_{33}^e r_{33}^h} \right\vert^2 + \left\vert \frac{t_{23}^h t_{31}^e}{1+r_{33}^e r_{33}^h} \right\vert^2 \ .
\end{equation}

\subsubsection{Case $\Gamma=0$}

For the case of total reflection at the Schottky barrier the conductance is given by
\begin{equation}
 g= \left\vert t_{21}^e+\frac{t_{23}^e  t_{31}^e}{1+r_{33}^e} \right\vert^2 \ .
\end{equation}
Clearly, there is no electron-hole scattering processes in this case.

\subsection{Conductance distribution}

In order to calculate the distribution of conductances we need to average over
the orthogonal or unitary ensembles ($\beta=1$ and $\beta=2$), as in the previous Section. See Eq.~(\ref{defpofg}). In the three-terminal configuration, however, the expression for the conductance [Eqs.~(\ref{gtwoterminal})-(\ref{s21eh})] is more complicated, even at the values of the transparency $\Gamma = 0$ and 1, where the expressions for the scattering-matrix elements $s_{21}^{ee}$ and $s_{21}^{eh}$ are simpler. Therefore we are not able to provide analytical expressions for $P(g)$, as for the two-terminal setup. Thus
we proceed with our analysis by generating numerically an ensemble of conductances from Eqs.~(\ref{gtwoterminal})-(\ref{definitions1}), where the elements of the scattering matrix associated to the quantum dot are generated according to the orthogonal and unitary ensembles.

\begin{figure}
\includegraphics[width=0.95\columnwidth]{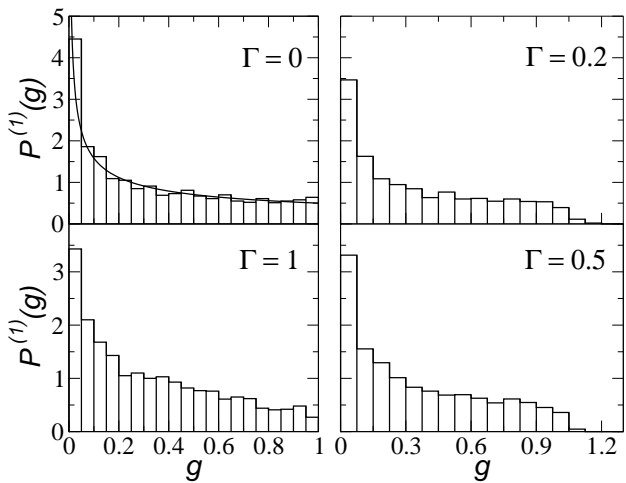}
\caption{Distribution of the conductance $P^{(1)}(g)$ for the three-terminal setup for different values of the transparency $\Gamma$. Notice that $P^{(1)}(g>1)>0$ for $0<\Gamma<1$, see right panels. The solid curve for $\Gamma=0$ is given by $P^{(1)}(g)=1/\sqrt{4g}$.}
\label{fig_6}
\end{figure}

In presence of time-reversal symmetry ($\beta=1$), the distribution of the conductance at different values of the transparency is shown in Fig.~\ref{fig_6}. At zero transparency, the distribution is described by\cite{baranger, jalabert} $P^{(1)}(g) =1/\sqrt{4g}$ (solid line in Fig.~\ref{fig_6}), which is the known result for a quantum-chaotic dot with two normal single-channel leads, i.e., there is no effect of the superconducting region on the conductance, as one might expect for a nontransparent Schottky barrier. For $\Gamma>0$, however, $P^{(1)}(g)$ is affected. From Fig.~\ref{fig_6} we can see that for intermediate values of the transparency ($\Gamma=0.2$ and $\Gamma=0.5$) the distribution probability is finite for $g > 1$, whereas for null  and complete transparency ($\Gamma =0$ and $\Gamma=1$, respectively) the probability is confined in the interval $0 < g < 1$. Therefore multiple scattering in the third lead, where the Schottky barrier is present,  produces larger values of conductance.

\begin{figure}
\includegraphics[width=0.95\columnwidth]{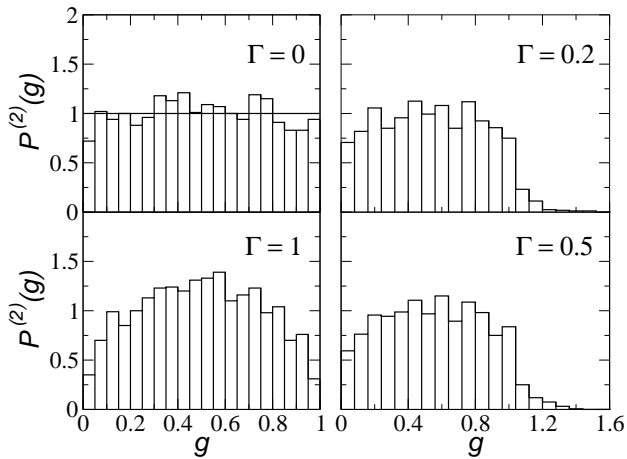}
\caption{Distribution of the conductance $P^{(2)}(g)$ for the three-terminal setup for different values of the transparency $\Gamma$. As for $\beta=1$, here $P^{(2)}(g>1)>0$ for $0<\Gamma<1$, see right panels. The horizontal line for $\Gamma=0$ corresponds to $P^{(2)}(g)=1$. }
\label{fig_7}
\end{figure}

When time reversal symmetry is broken ($\beta=2$) the distribution of the conductance is quite different from the $\beta=1$ case. In Fig.~\ref{fig_7} we show $P^{(2)}(g)$ at different values of the transparency. First we notice that for $\Gamma=0$, $P^{(2)}(g)$ reduces to the known result\cite{baranger,jalabert} $P^{(2)}(g)=1$ (shown as the solid line in Fig.~\ref{fig_7}). At intermediate values of the transparency of the Schottky barrier ($\Gamma=0.2$ and $\Gamma=0.5$), we observe again that multiple scattering in the superconducting lead can produce larger values of conductance than for the cases $\Gamma=0$ and $\Gamma=1$, i.e., the distribution $P^{(2)}(g)$ develops a tail for $g>1$ at intermediate values of $\Gamma$.

\section{summary and conclusions}

We have studied the effects of a superconducting lead attached to a normal
quantum-chaotic dot on the statistical properties of the conductance using random matrix theory and a scattering approach to the transport in normal-metal--superconducting (NS) hybrid structures.
We have calculated the complete distribution of the conductance in the presence
and absence of time-reversal symmetry ($\beta=1$ and $\beta=2$, respectively)
and for two different configurations: two and three-terminal setups. The conductance distribution has been analyzed as a function of the transparency $\Gamma$ of the Schottky barrier which is modeled as a planar barrier formed at the NS interface. The importance of considering the effect of a potential barrier formed at NS junctions has been already recognized in transport experiments,\cite{jespersen} however, opposite to the one open channel case that we study here, in these experiments a large number of channels are open.

For the two-terminal configuration we provided analytical expressions for the distribution of the conductance. We found that the distribution becomes more and more concentrated at $g=0$ as the value of the transparency is reduced, for both symmetries $\beta=1$ and $\beta=2$. This behavior might be expected considering that the conductance is due to the scattering at the NS interface (Andreev reflection), i.e., at zero transparency of the barrier the situation is reduced to a normal quantum dot with a normal lead attached where only reflection processes at the lead are present.

For the three-terminal configuration, the analytical expressions for the total scattering matrix and the conductance are more complicated than for the two-terminal case. We thus obtain the distribution of the conductance from an
ensemble of conductances generated numerically according to our analytical expressions for the elements of the total scattering matrix. We found that in presence and absence of  time-reversal symmetry, conductance values larger than unity are possible at intermediate values of the transparency of the Schottky barrier. As a consequence, the distribution of the conductance is broader than for the cases of zero transparency ($\Gamma=0$) and an ideal NS interface ($\Gamma=1$), i.e., multiple scattering in the superconducting region with the Schottky barrier increases the possibility of larger values of the conductance.

To conclude, we have seen that the presence of a Schottky barrier in the lead, which is formed in real experiments due to the mismatch of the conduction bands of the normal-metal and the superconductor, has a significant effect at the level of the distribution of the conductance. The progress in the fabrication of nanoscale heterostructures has made possible the experimental study of statistical properties of the conductance through such devices.\cite{jespersen} These experiments have pointed out the relevance of non-perfect contacts on the study of conductance. We thus think that the effects on the statistics of the conductance through quantum dots due to the presence of Andreev scattering and potential barriers at the leads, as those described here, are of experimental interest.

\acknowledgments

This work was supported by the ``Ram\'on y Cajal" program of the Spanish Ministry of Education and Science and Fondo Social Europeo.  J. A. M.-B. thanks the Depto. de F\'isica Te\'orica for its hospitality, as well as the program Ayudas para Estancias de Investigadores de Excelencia of the University of Zaragoza for financial support. He also acknowledges support from CONACyT Mexico (grant CB-2006-01-60879) and VIEP-BUAP. A. H. A. acknowledges hospitality at Depto. de F\'isica Te\'orica, Universidad de Zaragoza,  and financial support from the Gobierno de Arag\'on (Programa de Movilidad de Investigadores M1034-2007) and BIFI.

\end{document}